%% file: main.tex
\documentclass{lmcs}
\input{packages}

\input{macros}

\pdfoutput=1

\theoremstyle{plain}\newtheorem{rmrk}[thm]{Remark}

\begin{document}

\title[Fully graphical treatment of the quantum algorithm for the HSP]{Fully graphical treatment of the quantum algorithm for the Hidden Subgroup Problem}

\author[S. Gogioso]{Stefano Gogioso}
\address{Quantum Group, University of Oxford, UK}	

\author[A. Kissinger]{Aleks Kissinger} 
\address{iCIS, Radboud University. Nijmegen, Netherlands} 


\begin{abstract}
\noindent The abelian Hidden Subgroup Problem (HSP) is extremely general, and many problems with known quantum exponential speed-up (such as integers factorisation, the discrete logarithm and Simon's problem) can be seen as specific instances of it. The traditional presentation of the quantum protocol for the abelian HSP is low-level, and relies heavily on the the interplay between classical group theory and complex vector spaces. Instead, we give a high-level diagrammatic presentation which showcases the quantum structures truly at play. Specifically, we provide the first fully diagrammatic proof of correctness for the abelian HSP protocol, showing that strongly complementary observables are the key ingredient to its success. Being fully diagrammatic, our proof extends beyond the traditional case of finite-dimensional quantum theory: for example, we can use it to show that Simon's problem can be efficiently solved in real quantum theory, and to obtain a protocol that solves the HSP for certain infinite abelian groups.
\end{abstract}

\maketitle

\section{Introduction} 
\label{section_intro}

The advent of quantum computing promises to solve a number number of problems which have until now proven intractable for classical computers. Amongst these, one of the most famous is Shor's algorithm \cite{Shor1995,Ekert1996}: it allows for an efficient solution of the integer factorisation problem and the discrete logarithm problem, the hardness of which underlies many of the cryptographic algorithms which we currently entrust with our digital security (such as RSA and DHKE). Integer factorisation and the discrete logarithm, together with Simon's problem \cite{Simon1997}, Deutsch original algorithm and a number of other number-theoretic questions, turn out to be special cases of the much more general abelian Hidden Subgroup Problem (HSP) \cite{Jozsa2001}, and can be tackled by quantum computers using a uniform strategy.

The reformulation of Shor's algorithm as a special case of the abelian HSP \cite{Jozsa2001} makes the core issue of order-finding pop out as a group-theoretic question, and highlights the role played by the quantum Fourier transform in solving it \cite{Jozsa1997}. However, it is only with the compelling diagrammatic work of \cite{Vicary2012a} that the structures and information flow behind the quantum solution to the HSP become fully apparent: the unitary oracle used in the algorithm is decomposed into its algebraic building blocks, namely certain $\dagger$-Frobenius algebras, providing a clear topological account of why the procedure works. 

Strong complementarity \cite{Coecke2011}, a diagrammatic property of $\dagger$-Frobenius algebras related to the quantum Fourier transform, seems to play a fundamental role in the procedure. However, the diagrammatics in the proof of \cite{Vicary2012a} are nothing but straightforward graphical transcriptions of results obtained via traditional representation theory: this approach makes it hard to tell whether strong complementarity is truly fundamental at the topological level of information flow, or whether its presence is merely accidental. In this work, we provide the first fully diagrammatic proof of correctness for the quantum algorithm solving the abelian HSP, showing beyond any shadow of doubt that the feature providing the quantum advantage is indeed strong complementarity. Furthermore, our fully diagrammatic approach means that our results can be immediately generalised from traditional quantum mechanics to other theories featuring the necessary algebraic structures.

Both the original \cite{Vicary2012a} and this work use the graphical language of string diagrams and dagger symmetric monoidal categories \cite{Joyal1991,Joyal1996,Selinger2009,Coecke2009p,Kissinger2012,Coecke2016a}, where processes between quantum systems are represented by means of boxes and decorations (the processes) connected by wires (the input/output systems of sequentially composed processes) and/or stacked side by side (for parallel composition). This work adopts a left-to-right convention for sequential composition: this is in line with several other formalisms used in quantum information and computation, but is different from the bottom-to-top convention of \cite{Vicary2012a} and other works in string diagrams.

\section{The Hidden Subgroup Problem}
\label{section_HSP}

The \textbf{Hidden Subgroup Problem} (HSP) can be phrased as follows: 
\begin{enumerate}
\item[(i)] a finite group $G$ is fixed;
\item[(ii)] we are given an oracle implementing a \textbf{subgroup hiding function} $f : G \rightarrow \integersMod{2}^N$, which associates to each element of $G$ a \textbf{label} in the form of an $N$-bit string;
\item[(iii)] we are promised that the function is constant on (left) cosets of some subgroup $H \leq G$, and associates different labels to different cosets; equivalently, we are promised that $f$ factorizes as follows for some injective function $s$ and the quotient group homomorphism $q$ (we refer to this as the \textbf{factorisation promise}):
\begin{equation}\label{diagram_quotientFactorisation}
\input{pictures/quotientFactorisation.tikz}
\end{equation}
\item[(iv)] we are asked to find the \textbf{hidden subgroup} $H$.
\end{enumerate}
In the \textbf{abelian} HSP we are also promised that $G$ is abelian, while in the more general \textbf{normal} HSP we are promised that $H$ is a normal subgroup (a fact which always holds in the abelian HSP). In order for a quantum treatment to be possible at all, one imposes additional requirement on the oracle encoding the subgroup hiding function: 
\begin{enumerate}
	\item[(e)] the oracle is given \textit{coherently}, as the following unitary $U_f \in \UnitaryOps{\complexs[G] \otimes \complexs[\integersMod{2}^N]}$:
		\begin{equation}\label{eqn_unitaryOracle}
			U_f := \ket{g} \otimes \ket{t} \mapsto \ket{g} \otimes \ket{f(g) \oplus t}
		\end{equation}
	where by $\oplus$ we denoted the bit-wise XOR operation on $N$-bit strings.
\end{enumerate}

\noindent A number of important problems arise as special instances of the abelian HSP. In the Discrete Logarithm problem, one is given a prime number $p$, a primitive root $g$ mod $p$, and a number $a$ such that $a = \modclass{g^b}{p}$ for some unknown $b$ to be found. This is an instance of the abelian HSP with group $G = \integersMod{p-1} \times \integersMod{p-1}$, hidden subgroup $H = \integersMod{p-1} \cdot (b,1) \mathrel{\unlhd} G$ and subgroup hiding function $f(x,y) := \modclass{g^x a^{-y}}{p} = \modclass{g^{x-by}}{p}$.

In the Integer Factorisation problem, one is given a composite number $N$ and is asked to provide a non-trivial factorisation for it. Shor's algorithm solves the problem efficiently on quantum computers: its core is the order-finding subroutine, which considers an integer $a$ coprime\footnote{If $a \in \{2,...,N-1\}$ is not coprime with $N$, then we already have a non-trivial factorisation of $N$.} with $N$, and asks for the order of $a$ as a multiplicative unit modulo $N$. The order-finding subroutine is an instance of the abelian HSP with group $G = \integersMod{N}^\times$ (the abelian group of multiplicative units modulo $N$)\footnote{In practice one uses $\integersMod{2^M}$: if $M \gg \log_2 N$, the errors due to the inexact period of $a$ will be small.}, hidden subgroup $H = \langle a \rangle \isom \integersMod{\operatorname{ord}(a)}$ and subgroup hiding function $f(x) = \modclass{a^x}{N}$.

In Simon's problem, one is given a function $f : \integersMod{2}^N \rightarrow \integersMod{2}^N$ with the promise that the stabilizer subgroup for $f$ has order $2$: there is a unique non-zero string $z \in \integersMod{2}^N$, which we are asked to find, such that for any two $N$-bit strings $x,y \in \integersMod{2}^N$ we have that $f(x) = f(y)$ if and only if $x = y$ or $x = y \oplus z$. The importance of Simon's problem in the  complexity of quantum computing lies in a result \cite{Bernstein1997} stating that, relative to oracles with the promise above, Simon's problem separates BQP (the class of bounded-error quantum polynomial time problems) from BPP (the class of bounded-error classical polynomial time problems). Simon's problem is clearly an instance of the abelian HSP, with $G = \integersMod{2}^N$, hidden subgroup $H = \langle z \rangle = \{0,z\}$ and subgroup hiding function $f$.

Quantum algorithms to solve the HSP have been studied beyond the abelian case. An extension of the efficient quantum solution to the case of normal subgroups of non-abelian groups is given by \cite{Hallgren2000}, while \cite{Moore2008,Hallgren2010} provide a no-go theorem showing that the same techniques cannot be used to formulate an efficient quantum solution to the general non-abelian case. The general non-abelian case is important because two interesting problems of classical computational complexity arise as special cases: the Graph Isomorphism Problem arises as a special case of the HSP on symmetric groups \cite{Hallgren2000}, while the Unique Shortest Vector Problem (uSVP) arises as a special case of the HSP on dihedral groups \cite{Regev2004}. The latter forms the basis of a public key cryptosystem \cite{Regev2004a} which, subject to quantum intractability of the HSP on dihedral groups, is a candidate to replace RSA in post-quantum cryptography (as are many other lattice-based cryptographic algorithms).

\section{Coherent Data Manipulation}
\label{section_CoherentOps}

At the basis of the diagrammatic treatment of the HSP is the observation that the unitary oracle of Equation \ref{eqn_unitaryOracle} has certain Frobenius algebras as its constituent parts \cite{Vicary2012a}. Frobenius algebras, or more precisely $\dagger$-Frobenius algebras, are a fundamental structure in the categorical and diagrammatic treatment of quantum information and computation: they provide the coherent versions of several classical data manipulation primitives, such as copy, deletion, matching and group operations (e.g. bit-wise XOR, modular addition, etc), they obey simple graphical rules, and they can be composed together to form more complex quantum processes taking place during quantum computation. Our main contribution will be to show that those abstract graphical rules are in fact sufficient to prove correctness of the quantum subroutine for the HSP. 

We begin our journey by observing that there are four distinct group structures playing a role in the definition of the Hidden Subgroup Problem, and/or intervening in some capacity in the formulation of the quantum algorithm aiming to solve it:
\begin{enumerate} 
	\item[(a)] the main group $G$;
	\item[(b)] the hidden subgroup $H$;
	\item[(c)] the quotient group $G/H$ appearing in the factorisation promise;
	\item[(d)] the group $\integersMod{2}^N$ of $N$-bit strings under the bit-wise XOR operation.
\end{enumerate}
Two distinct algebraic structures intervene when we wish to encode a finite group $K$ (such as $K=G,H,G/H,\integersMod{2}^N$ above) into a quantum system $\SpaceH$: 
\begin{enumerate}
	\item[(a)] a \textbf{copy/delete coalgebra}  $(\!\hbox{\input{symbols/ZbwcomultSym.tex}}\!\!_K, \!\hbox{\input{symbols/ZbwcounitSym.tex}}\!\!_K)$ on $\SpaceH$, defining a coherent encoding of the elements (or \textit{points}) of the group as an orthonormal basis of the quantum system;
	\item[(b)] a \textbf{group algebra} $(\!\hbox{\input{symbols/DmultSym.tex}}\!\!_K, \!\hbox{\input{symbols/DunitSym.tex}}\!\!_K)$ on $\SpaceH$, defining the group multiplication and unit on the coherently encoded points.
\end{enumerate}
Both structures will turn out to define certain $\dagger$-Frobenius algebras, corresponding in turn to various quantum observables intervening in algorithm itself.

\subsection{The copy/delete coalgebra}

The copy/delete coalgebra $(\!\hbox{\input{symbols/ZbwcomultSym.tex}}\!\!_K, \!\hbox{\input{symbols/ZbwcounitSym.tex}}\!\!_K)$ defines the coherent copy and delete operations on the encoded points:
\begin{equation}\label{coherentCopyDelete}
\input{pictures/coherentCopyDelete.tikz}
\end{equation}
The adjoints of the linear maps in \ref{coherentCopyDelete} form an algebra $(\!\hbox{\input{symbols/ZbwmultSym.tex}}\!\!_K, \!\hbox{\input{symbols/ZbwunitSym.tex}}\!\!_K)$ on $\SpaceH$, defining the coherent match operation and the superposition of the encoded classical data:
\begin{equation}\label{coherentMatchSuperposition}
\input{pictures/coherentMatchSuperposition.tikz}
\end{equation}
The algebra and coalgebra are not unrelated, but instead interact via the following Frobenius law, which can be interpreted as encoding a basic topological property of the flow of classical information:
\begin{equation}\label{FrobeniusLaw}
\input{pictures/FrobeniusLaw.tikz}
\end{equation}
As a consequence, the four coherent operations defined above form a $\dagger$-Frobenius algebra, which we will refer to as the \textbf{point structure} and denote by $\hbox{\input{symbols/ZbwdotSym.tex}}\!\!_K$ for short. In fact, they satisfy the additional requirements of commutativity and speciality below, and hence form a special commutative $\dagger$-Frobenius algebra ($\dagger$-SCFA for short):
\begin{equation}\label{specialCommutativeLaws}
\input{pictures/specialCommutativeLaws.tikz}
\end{equation}
\noindent A key result of categorical quantum mechanics \cite{Coecke2013b} states that there is an exact correspondence between $\dagger$-SCFAs on a finite-dimensional quantum system $\SpaceH$ and orthonormal bases (i.e. non-degenerate quantum observables) of $\SpaceH$: all $\dagger$-SCFAs take the form of a coherent copy/delete coalgebra and a match/superimpose algebra for some orthonormal basis. When we said that the copy/delete coalgebra defines the coherent encoding of points, we meant this: the coalgebra defines its adjoint, and together they form a $\dagger$-SCFA, which in turns identifies a unique orthonormal basis.

If $\hbox{\input{symbols/ZbwdotSym.tex}}\!\!_A$ and $\hbox{\input{symbols/ZbwdotSym.tex}}\!\!_B$ are two $\dagger$-SCFAs, encoding classical input and output data respectively, the following diagrammatic equations \cite{Coecke2010} provide an exact characterisation of those complex linear maps arising as coherent versions (i.e. linear extensions) of classical maps\footnote{If the $\dagger$-SCFA $\hbox{\input{symbols/ZbwdotSym.tex}}\!\!_A$ is associated with an orthonormal basis $\big(\ket{a}\big)_{a \in A}$ of some space $\SpaceH$ and the $\dagger$-SCFA $\hbox{\input{symbols/ZbwdotSym.tex}}\!\!_B$ is associated with an orthonormal basis $\big(\ket{b}\big)_{b \in B}$ of some space $\SpaceK$, then a linear map $F: \SpaceH \rightarrow \SpaceK$ satisfy the two equations above if and only if we have $F\ket{a} = \ket{f(a)}$ for some classical map $f: X \rightarrow Y$.}:
\begin{equation}\label{eqns_classicalMap}
\input{pictures/classicalMap.tikz}
\end{equation}
If $F$ satisfies the three equations above, we will say that $F$ is \textbf{$\hbox{\input{symbols/ZbwdotSym.tex}}\!\!_A$-to-$\hbox{\input{symbols/ZbwdotSym.tex}}\!\!_B$ classical}. As a special case, we say that a state $\ket{\psi}: \complexs \rightarrow \SpaceH$ is \textbf{$\hbox{\input{symbols/ZbwdotSym.tex}}\!\!$-classical} if it is coherently copied, deleted and transposed by the $\dagger$-SCFA $\hbox{\input{symbols/ZbwdotSym.tex}}\!\!$ in the following sense:
\begin{equation}\label{eqns_classicalState}
\input{pictures/classicalState.tikz}
\end{equation}
In the case of finite-dimensional Hilbert spaces, classical states for a $\dagger$-SCFA $\hbox{\input{symbols/ZbwdotSym.tex}}\!\!$ are exactly the states in the associated orthonormal basis.

The graphical definitions of classical states (Equations \ref{eqns_classicalState}) and classical maps (Equations \ref{eqns_classicalMap}) make sense for arbitrary $\dagger$-Frobenius algebras in arbitrary dagger symmetric monoidal categories, and we extend them accordingly. Even though the exact correspondence with encoded classical data is lost, the classical states $\classicalStates{\hbox{\input{symbols/ZbwdotSym.tex}}\!\!}$ for a $\dagger$-Frobenius algebra $\hbox{\input{symbols/ZbwdotSym.tex}}\!\!$ can still be copied and deleted as if they were classical information, and the $\hbox{\input{symbols/ZbwdotSym.tex}}\!\!_A$-to-$\hbox{\input{symbols/ZbwdotSym.tex}}\!\!_B$ classical maps always restrict to functions $\classicalStates{\hbox{\input{symbols/ZbwdotSym.tex}}\!\!_A} \rightarrow \classicalStates{\hbox{\input{symbols/ZbwdotSym.tex}}\!\!_B}$ on the sets of classical states. We can also recover the notion of basis in this more general setting, by saying that a $\dagger$-Frobenius algebra $\hbox{\input{symbols/ZbwdotSym.tex}}\!\!$ has \textbf{enough classical states} if any two parallel maps coincide whenever they agree on all $\hbox{\input{symbols/ZbwdotSym.tex}}\!\!$-classical states.

\subsection{The group algebra}

The group algebra $(\!\hbox{\input{symbols/DmultSym.tex}}\!\!_K, \!\hbox{\input{symbols/DunitSym.tex}}\!\!_K)$ defines the coherent group multiplication on the encoded points, and labels one of them as the group unit\footnote{We use multiplicative notation for the groups operations, to accommodate the general case where the group in consideration might not be abelian. As a special exception, we will use additive notation $(\integersMod{2}^N,\oplus,0)$ for the bit-wise XOR group operation on $N$-bit strings, with the zero string as its unit.}:
\begin{equation}\label{groupAlgebra}
\input{pictures/groupAlgebra.tikz}
\end{equation} 
The explicit presentation given above is easy to understand in the traditional Hilbert space formalism, but unsatisfactory from our more abstract point of view. We would like to replace it with a series of diagrammatic laws, capturing the same structure while highlighting the interactions between the group algebra and the point structure. As we shall shortly see, the notion of strong complementarity is exactly what we are looking for.

Just like we did with the copy/delete coalgebra, we can take the adjoints of the maps in the group algebra to obtain a coalgebra $(\!\hbox{\input{symbols/DcomultSym.tex}}\!\!_K,\!\hbox{\input{symbols/DcounitSym.tex}}\!\!_K)$:
\begin{equation}
\input{pictures/groupAlgebraAdjoints.tikz}
\end{equation} 
Somewhat surprisingly, the four maps $(\!\hbox{\input{symbols/DmultSym.tex}}\!\!_K,\!\hbox{\input{symbols/DunitSym.tex}}\!\!_K,\!\hbox{\input{symbols/DcomultSym.tex}}\!\!_K,\!\hbox{\input{symbols/DcounitSym.tex}}\!\!_K)$ also define a $\dagger$-Frobenius algebra, which we shall refer to as the \textbf{group structure} and denote by $\hbox{\input{symbols/DdotSym.tex}}\!\!_K$. 

Contrary to the point structure, the group structure does not in general satisfy the speciality law, and it is commutative if and only if the group $K$ is. However, it is always quasi-special and balanced-symmetric, i.e. it satisfies the following laws, where $\xi$ is some invertible scalar (we refer to it as the \textbf{normalisation scalar}):
\begin{equation}\label{quasispecialBalancedSymmetricLaws}
\input{pictures/quasispecialBalancedSymmetricLaws.tikz}
\end{equation}
In the Hilbert space case explicitly defined by \ref{groupAlgebra}, the scalar occurring in the quasi-speciality law is $\xi:= \sqrt{|K|}$. The result of \cite{Coecke2013b} on orthonormal bases straightforwardly extends to quasi-special commutative $\dagger$-Frobenius algebras ($\dagger$-qSCFA for short), which correspond exactly to orthogonal bases where all elements have the same square norm (equal to $\xi^\dagger \xi$). Also, by the definition of the quasi-special law we have that every quasi-special $\dagger$-Frobenius algebra can be normalised to obtain a unique corresponding special $\dagger$-Frobenius algebra, and hence quasi-speciality is merely a diagrammatic convenience. An extension of the result of \cite{Coecke2013b} to arbitrary balanced-symmetric special\footnote{And hence also to balanced-symmetric quasi-special $\dagger$-Frobenius algebras.} $\dagger$-Frobenius algebra (balanced-symmetric $\dagger$-SFAs, for short) is obtained straightforwardly from \cite{Vicary2011}, and it states that balanced-symmetric $\dagger$-Frobenius algebras on finite-dimensional quantum system are in exact correspondence with complete families of orthogonal projectors (i.e. quantum observables).

\subsection{Strong complementarity} 

With these observations at hand, we are now ready to define the diagrammatic property of strong complementarity, and to show that it captures exactly the abstract relationship between the point and group structures defined explicitly above. We will state it in arbitrary dagger symmetric monoidal categories ($\dagger$-SMC for short).
\begin{defi}\label{def_complementarity}
	Two balanced-symmetric $\dagger$-qSFAs $\hbox{\input{symbols/ZbwdotSym.tex}}\!\!$ and $\hbox{\input{symbols/DdotSym.tex}}\!\!$ on the same object $\SpaceH$ are said to be \textbf{complementary} (or a \textbf{complementary pair}) if they satisfy the \textbf{Hopf Law}:
	\begin{equation}\label{hopfsLaw}
		\input{pictures/HopfsLaw.tikz}
	\end{equation}
	where the \textbf{antipode} $\hbox{\input{symbols/antipodeSym.tex}}\!: \SpaceH \rightarrow \SpaceH$ satisfies the following equation:
	\begin{equation}\label{antipode}
		\input{pictures/antipode.tikz}
	\end{equation}
	Note that the antipode is unitary (by definition) and self-adjoint (by Equation \ref{antipode}).
\end{defi}

\begin{rmrk}
Because of balanced-symmetry and the self-adjointness requirement, the antipode can be equivalently written in two additional ways:
	\begin{equation}\label{antipodeConj}
		\input{pictures/antipodeConj.tikz}
	\end{equation}
\end{rmrk}
\begin{defi}\label{def_strongComplementarity}
	Two balanced-symmetric $\dagger$-qSFAs $\hbox{\input{symbols/ZbwdotSym.tex}}\!\!$ and $\hbox{\input{symbols/DdotSym.tex}}\!\!$ on the same object of a $\dagger$-SMC are said to be \textbf{strongly complementary} (or a \textbf{strongly complementary pair}) if they are complementary and furthermore satisfy the following four equations:
	\begin{equation}\label{eqns_strongComplementarity}
	\input{pictures/strongComplementarity4.tikz}
	\end{equation}
	The empty diagram on the right hand side of the top-right equation corresponds to the scalar 1, i.e. the identity morphism $1_I$ on the monoidal unit $I$.
\end{defi}
\begin{thm}\label{thm_StrongComplementarity}
Let $(\hbox{\input{symbols/ZbwdotSym.tex}}\!\!, \hbox{\input{symbols/DdotSym.tex}}\!\!)$ be a pair of balanced-symmetric $\dagger$-qSFAs. If $(\hbox{\input{symbols/ZbwdotSym.tex}}\!\!, \hbox{\input{symbols/DdotSym.tex}}\!\!)$ is a strongly complementary pair then the algebraic fragment of $\hbox{\input{symbols/DdotSym.tex}}\!\!$ endows the $\hbox{\input{symbols/ZbwdotSym.tex}}\!\!$-classical states with the structure of a group $(\classicalStates{\hbox{\input{symbols/ZbwdotSym.tex}}\!\!},\!\hbox{\input{symbols/DmultSym.tex}}\!\!\!,\!\hbox{\input{symbols/DunitSym.tex}}\!\!\!)$, which has the antipode $\hbox{\input{symbols/antipodeSym.tex}}\!$ as group inverse and is abelian if and only if $\hbox{\input{symbols/DdotSym.tex}}\!\!$ is commutative. A converse holds when $\hbox{\input{symbols/ZbwdotSym.tex}}\!\!$ has enough classical states and $\!\hbox{\input{symbols/ZbwunitSym.tex}}\!\!$ is a $\hbox{\input{symbols/DdotSym.tex}}\!\!$-classical state\footnote{In fact it is sufficient for $\!\hbox{\input{symbols/ZbwunitSym.tex}}\!\!$ to satisfy the $\hbox{\input{symbols/DdotSym.tex}}\!\!$ transpose condition. The reasons behind this somewhat bizarre requirement are explained after the proof.}: in that case, if $(\classicalStates{\hbox{\input{symbols/ZbwdotSym.tex}}\!\!},\!\hbox{\input{symbols/DmultSym.tex}}\!\!\!,\!\hbox{\input{symbols/DunitSym.tex}}\!\!\!)$ is a group then $(\hbox{\input{symbols/ZbwdotSym.tex}}\!\!, \hbox{\input{symbols/DdotSym.tex}}\!\!)$ is a strongly complementary pair.
\end{thm}
\proof
A first version of this result was proven in \cite{Coecke2011,Kissinger2012} for the finite abelian group case of $\dagger$-SCFAs on finite-dimensional Hilbert spaces (which always have enough points). Here we present a simpler proof of much broader scope, valid for all balanced-symmetric $\dagger$-qSFAs in arbitrary $\dagger$-SMCs.

\noindent To begin with, assume that $(\hbox{\input{symbols/ZbwdotSym.tex}}\!\!, \hbox{\input{symbols/DdotSym.tex}}\!\!)$ is a strongly complementary pair. The defining Equations \ref{eqns_strongComplementarity} of strong complementarity state that the unit $\!\hbox{\input{symbols/DunitSym.tex}}\!\!$ and multiplication $\!\hbox{\input{symbols/DmultSym.tex}}\!\!$ satisfy the copy and delete conditions for $\hbox{\input{symbols/ZbwdotSym.tex}}\!\!$-classical states and $(\hbox{\input{symbols/ZbwdotSym.tex}}\!\! \otimes \hbox{\input{symbols/ZbwdotSym.tex}}\!\!)$-to-$\hbox{\input{symbols/ZbwdotSym.tex}}\!\!$ classical maps. In order to prove that they are indeed classical (so that $(\classicalStates{\hbox{\input{symbols/ZbwdotSym.tex}}\!\!},\!\hbox{\input{symbols/DmultSym.tex}}\!\!,\!\hbox{\input{symbols/DunitSym.tex}}\!\!)$ is a monoid) we need to show the transpose condition for both:
\begin{equation}\label{strongComplementarityAdjoinConditions}
	\input{pictures/strongComplementarityAdjoinConditions.tikz}
\end{equation}
Both equations can be shown to follow from Hopf's law, the self-adjointness requirement for the antipode, and the four defining equations of strong complementarity:
\begin{equation}\label{HopfsLawClassicalAdjunctionsConsequence}
	\resizebox{\textwidth}{!}{\input{pictures/HopfsLawClassicalAdjunctionsConsequence.tikz}}
\end{equation}
\begin{equation}\label{HopfsLawClassicalAdjunctionsConsequence2}
	\resizebox{\textwidth}{!}{\input{pictures/HopfsLawClassicalAdjunctionsConsequence2.tikz}}
\end{equation}
Finally, applying Hopf's law to any $\hbox{\input{symbols/ZbwdotSym.tex}}\!\!$-classical state (which is copied by $\!\hbox{\input{symbols/ZbwcomultSym.tex}}\!\!$ and deleted by $\!\hbox{\input{symbols/ZbwcounitSym.tex}}\!\!$) shows that the antipode $\hbox{\input{symbols/antipodeSym.tex}}\!$ acts as an inverse for elements of the the monoid $(\classicalStates{\hbox{\input{symbols/ZbwdotSym.tex}}\!\!},\!\hbox{\input{symbols/DmultSym.tex}}\!\!,\!\hbox{\input{symbols/DunitSym.tex}}\!\!)$, which is therefore a group. As a side note, a proof on the same lines of that in \ref{HopfsLawClassicalAdjunctionsConsequence}, but with colours swapped, shows that $\!\hbox{\input{symbols/ZbwunitSym.tex}}\!\!$ satisfies the transpose condition with respect to $\hbox{\input{symbols/DdotSym.tex}}\!\!$, and hence is always a $\hbox{\input{symbols/DdotSym.tex}}\!\!$-classical state.

Conversely, assume that $\hbox{\input{symbols/ZbwdotSym.tex}}\!\!$ has enough classical states, and that $(\classicalStates{\hbox{\input{symbols/ZbwdotSym.tex}}\!\!},\!\hbox{\input{symbols/DmultSym.tex}}\!\!,\!\hbox{\input{symbols/DunitSym.tex}}\!\!)$ is a group. In particular, $\!\hbox{\input{symbols/DunitSym.tex}}\!\!$ is a $\hbox{\input{symbols/ZbwdotSym.tex}}\!\!$-classical state and $\!\hbox{\input{symbols/DmultSym.tex}}\!\!$ is $(\hbox{\input{symbols/ZbwdotSym.tex}}\!\! \otimes \hbox{\input{symbols/ZbwdotSym.tex}}\!\!)$-to-$\hbox{\input{symbols/ZbwdotSym.tex}}\!\!$-classical map: because $\hbox{\input{symbols/ZbwdotSym.tex}}\!\!$ has enough classical states, the four defining Equations \ref{eqns_strongComplementarity} of strong complementarity and the two Equations \ref{strongComplementarityAdjoinConditions} all hold. All that remains to show is that $(\hbox{\input{symbols/ZbwdotSym.tex}}\!\!,\hbox{\input{symbols/DdotSym.tex}}\!\!)$ are complementary. The proof of Hopf's law goes as follows, and requires one to use the fact that $\!\hbox{\input{symbols/ZbwunitSym.tex}}\!\!$ is transposed by $\hbox{\input{symbols/DdotSym.tex}}\!\!$:
\begin{equation}\label{eqn_HopfProof}
\resizebox{\textwidth}{!}{\input{pictures/HopfLawProof.tikz}}
\end{equation}
The proof for the other equation is similar (using balanced-symmetry at one point). Finally, Hopf's law proves that the antipode acts as the group inverse on the $\hbox{\input{symbols/ZbwdotSym.tex}}\!\!$-classical states: because $\hbox{\input{symbols/ZbwdotSym.tex}}\!\!$ has enough classical states, the antipode must be self-inverse, or equivalently self-adjoint, completing the proof of complementarity. 
\qed

The requirement in Theorem \ref{thm_StrongComplementarity} that $\!\hbox{\input{symbols/ZbwunitSym.tex}}\!\!$ be $\hbox{\input{symbols/DdotSym.tex}}\!\!$-classical\footnote{Or at least transposed by $\hbox{\input{symbols/DdotSym.tex}}\!\!$, since the copy and delete conditions for $\!\hbox{\input{symbols/ZbwunitSym.tex}}\!\!$ to be $\hbox{\input{symbols/DdotSym.tex}}\!\!$-classical can already be obtained as the adjoints of the delete conditions for $\!\hbox{\input{symbols/DunitSym.tex}}\!\!$ and $\!\hbox{\input{symbols/DmultSym.tex}}\!\!$.} 
seems to come somewhat out of the blue, but is instead an inkling of a deeper result. Theorem \ref{thm_StrongComplementarity} was stated asymmetrically and with a minimal set of requirements and conclusions, which are relevant when $\hbox{\input{symbols/ZbwdotSym.tex}}\!\!$ and $\hbox{\input{symbols/DdotSym.tex}}\!\!$ are the point and group structures for some group $K$. However, the definitions of complementarity and strong complementarity are fully symmetric in the two $\dagger$-Frobenius algebras involved, and Theorem \ref{thm_StrongComplementarity} can be re-formulated in a fully symmetric way.
\begin{cor}\label{cor_StrongComplementaritySymmetric}
Let $(\hbox{\input{symbols/ZbwdotSym.tex}}\!\!, \hbox{\input{symbols/DdotSym.tex}}\!\!)$ be a pair of balanced-symmetric $\dagger$-qSFAs. If $(\hbox{\input{symbols/ZbwdotSym.tex}}\!\!, \hbox{\input{symbols/DdotSym.tex}}\!\!)$ is a strongly complementary pair then both $(\classicalStates{\hbox{\input{symbols/ZbwdotSym.tex}}\!\!},\!\hbox{\input{symbols/DmultSym.tex}}\!\!\!,\!\hbox{\input{symbols/DunitSym.tex}}\!\!\!)$ and $(\classicalStates{\hbox{\input{symbols/DdotSym.tex}}\!\!},\!\hbox{\input{symbols/ZbwmultSym.tex}}\!\!\!,\!\hbox{\input{symbols/ZbwunitSym.tex}}\!\!\!)$ are groups, both with the antipode as group inverse. A converse holds when at least one of $(\hbox{\input{symbols/ZbwdotSym.tex}}\!\!,\hbox{\input{symbols/DdotSym.tex}}\!\!)$ has enough classical states: in that case, if $(\classicalStates{\hbox{\input{symbols/ZbwdotSym.tex}}\!\!},\!\hbox{\input{symbols/DmultSym.tex}}\!\!\!,\!\hbox{\input{symbols/DunitSym.tex}}\!\!\!)$ and $(\classicalStates{\hbox{\input{symbols/DdotSym.tex}}\!\!},\!\hbox{\input{symbols/ZbwmultSym.tex}}\!\!\!,\!\hbox{\input{symbols/ZbwunitSym.tex}}\!\!\!)$ are both groups then $(\hbox{\input{symbols/ZbwdotSym.tex}}\!\!, \hbox{\input{symbols/DdotSym.tex}}\!\!)$ is a strongly complementary pair. \qed
\end{cor}

Having just seen the correspondence between groups and strongly complementary pairs, it is natural to ask how group homomorphisms could be characterised diagrammatically. 
\begin{defi}
If $(\hbox{\input{symbols/ZbwdotSym.tex}}\!\!_H,\hbox{\input{symbols/DdotSym.tex}}\!\!_H)$ and $(\hbox{\input{symbols/ZbwdotSym.tex}}\!\!_K,\hbox{\input{symbols/DdotSym.tex}}\!\!_K)$ are strongly complementary pairs on two objects $\SpaceH$ and $\SpaceK$ respectively, we say that a process $F: \SpaceH \rightarrow \SpaceK$ is a \textbf{$(\hbox{\input{symbols/ZbwdotSym.tex}}\!\!_H,\hbox{\input{symbols/DdotSym.tex}}\!\!_H)$-to-$(\hbox{\input{symbols/ZbwdotSym.tex}}\!\!_K,\hbox{\input{symbols/DdotSym.tex}}\!\!_K)$ homomorphism} if it is $\hbox{\input{symbols/ZbwdotSym.tex}}\!\!_H$-to-$\hbox{\input{symbols/ZbwdotSym.tex}}\!\!_K$ classical and furthermore satisfies the following\footnote{Note that the inverse condition follows from the other two when $\hbox{\input{symbols/ZbwdotSym.tex}}\!\!_H$ has enough classical states.}:
\begin{equation}\label{groupHomom}
\input{pictures/groupHomom.tikz}
\end{equation}
\end{defi}
\begin{lem}\label{lem_groupHomom}
Let $(\hbox{\input{symbols/ZbwdotSym.tex}}\!\!_H,\hbox{\input{symbols/DdotSym.tex}}\!\!_H)$ and $(\hbox{\input{symbols/ZbwdotSym.tex}}\!\!_K,\hbox{\input{symbols/DdotSym.tex}}\!\!_K)$ be two strongly complementary pairs on two objects $\SpaceH$ and $\SpaceK$ respectively. If $F: \SpaceH \rightarrow \SpaceK$ is a $(\hbox{\input{symbols/ZbwdotSym.tex}}\!\!_H,\hbox{\input{symbols/DdotSym.tex}}\!\!_H)$-to-$(\hbox{\input{symbols/ZbwdotSym.tex}}\!\!_K,\hbox{\input{symbols/DdotSym.tex}}\!\!_K)$ homomorphism, then restricting $F$ to the $\hbox{\input{symbols/ZbwdotSym.tex}}\!\!_H$-classical states gives rise to a well-defined group homomorphism $(\classicalStates{\hbox{\input{symbols/ZbwdotSym.tex}}\!\!_H},\!\hbox{\input{symbols/DmultSym.tex}}\!\!_H,\!\hbox{\input{symbols/DunitSym.tex}}\!\!_H) \rightarrow (\classicalStates{\hbox{\input{symbols/ZbwdotSym.tex}}\!\!_K},\!\hbox{\input{symbols/DmultSym.tex}}\!\!_K,\!\hbox{\input{symbols/DunitSym.tex}}\!\!_K)$. A converse to this statement holds if $\hbox{\input{symbols/ZbwdotSym.tex}}\!\!_H$ has enough classical states: in that case, any $F:\SpaceH \rightarrow \SpaceK$ which gives rise to a well-defined group homomorphism  $(\classicalStates{\hbox{\input{symbols/ZbwdotSym.tex}}\!\!_H},\!\hbox{\input{symbols/DmultSym.tex}}\!\!_H,\!\hbox{\input{symbols/DunitSym.tex}}\!\!_H) \rightarrow (\classicalStates{\hbox{\input{symbols/ZbwdotSym.tex}}\!\!_K},\!\hbox{\input{symbols/DmultSym.tex}}\!\!_K,\!\hbox{\input{symbols/DunitSym.tex}}\!\!_K)$ when restricted to the $\hbox{\input{symbols/ZbwdotSym.tex}}\!\!_H$-classical states is in fact a  $(\hbox{\input{symbols/ZbwdotSym.tex}}\!\!_H,\hbox{\input{symbols/DdotSym.tex}}\!\!_H)$-to-$(\hbox{\input{symbols/ZbwdotSym.tex}}\!\!_K,\hbox{\input{symbols/DdotSym.tex}}\!\!_K)$ homomorphism.
\end{lem}
\proof
If $F$ is a $(\hbox{\input{symbols/ZbwdotSym.tex}}\!\!_H,\hbox{\input{symbols/DdotSym.tex}}\!\!_H)$-to-$(\hbox{\input{symbols/ZbwdotSym.tex}}\!\!_K,\hbox{\input{symbols/DdotSym.tex}}\!\!_K)$ homomorphism, then in particular it restricts to a function $\classicalStates{\hbox{\input{symbols/ZbwdotSym.tex}}\!\!_H} \rightarrow \classicalStates{\hbox{\input{symbols/ZbwdotSym.tex}}\!\!_K}$; then the two Equations \ref{groupHomom} state applied to $\hbox{\input{symbols/ZbwdotSym.tex}}\!\!_H$-classical states simply state that the function respects the group multiplication and unit, i.e. that it is a group homomorphism. 
Conversely, we have already seen that if $F$ restricts to a function then it is $\hbox{\input{symbols/ZbwdotSym.tex}}\!\!_H$-to-$\hbox{\input{symbols/ZbwdotSym.tex}}\!\!_K$ classical, since $\hbox{\input{symbols/ZbwdotSym.tex}}\!\!_H$ has enough classical states. Furthermore, Equations \ref{groupHomom} hold when applied to $\hbox{\input{symbols/ZbwdotSym.tex}}\!\!_H$-classical states because $F$ gives rise to a group homomorphism, and hence hold in full generality because $\hbox{\input{symbols/ZbwdotSym.tex}}\!\!_H$ has enough classical states. We conclude that $F$ is a $(\hbox{\input{symbols/ZbwdotSym.tex}}\!\!_H,\hbox{\input{symbols/DdotSym.tex}}\!\!_H)$-to-$(\hbox{\input{symbols/ZbwdotSym.tex}}\!\!_K,\hbox{\input{symbols/DdotSym.tex}}\!\!_K)$ homomorphism.
\qed

\begin{lem}
Let $(\hbox{\input{symbols/ZbwdotSym.tex}}\!\!_H,\hbox{\input{symbols/DdotSym.tex}}\!\!_H)$ and $(\hbox{\input{symbols/ZbwdotSym.tex}}\!\!_K,\hbox{\input{symbols/DdotSym.tex}}\!\!_K)$ be two strongly complementary pairs on two objects $\SpaceH$ and $\SpaceK$ respectively. Then a process $F:\SpaceH \rightarrow \SpaceK$ is a  $(\hbox{\input{symbols/ZbwdotSym.tex}}\!\!_H,\hbox{\input{symbols/DdotSym.tex}}\!\!_H)$-to-$(\hbox{\input{symbols/ZbwdotSym.tex}}\!\!_K,\hbox{\input{symbols/DdotSym.tex}}\!\!_K)$ homomorphism if and only if its adjoint $F^\dagger:\SpaceK \rightarrow \SpaceH$ is a $(\hbox{\input{symbols/DdotSym.tex}}\!\!_K,\hbox{\input{symbols/ZbwdotSym.tex}}\!\!_K)$-to-$(\hbox{\input{symbols/DdotSym.tex}}\!\!_H,\hbox{\input{symbols/ZbwdotSym.tex}}\!\!_H)$ homomorphism\footnote{Note that not only $H$ and $K$ were swapped, but that colours were swapped as well.}. 
\end{lem}
\proof
Being a homomorphism is a statement comprised of six conditions: the copy,delete and transpose conditions from Equations \ref{eqns_classicalMap} and the multiplication, unit and inverse conditions from Equations \ref{groupHomom}. Taking the dagger of the copy/delete conditions from $F$ being a $(\hbox{\input{symbols/ZbwdotSym.tex}}\!\!_H,\hbox{\input{symbols/DdotSym.tex}}\!\!_H)$-to-$(\hbox{\input{symbols/ZbwdotSym.tex}}\!\!_K,\hbox{\input{symbols/DdotSym.tex}}\!\!_K)$ homomorphism are exactly the multiplication/unit conditions for $F^\dagger$ being a $(\hbox{\input{symbols/DdotSym.tex}}\!\!_K,\hbox{\input{symbols/ZbwdotSym.tex}}\!\!_K)$-to-$(\hbox{\input{symbols/DdotSym.tex}}\!\!_H,\hbox{\input{symbols/ZbwdotSym.tex}}\!\!_H)$ homomorphism; conversely, taking the dagger of the copy/delete conditions for $F^\dagger$ yields the multiplication/unit conditions for $F$. Also, taking the dagger of the inverse condition for $F$ yields the inverse condition for $F^\dagger$, as the antipodes are self-adjoint. Finally, we need to show that the transpose condition for $F$ implies the transpose condition for $F^\dagger$. To obtain the desired implication, we pre/post compose the transpose condition for $F$ with the symmetric cup and cap for $\hbox{\input{symbols/DdotSym.tex}}\!\!_H$ and $\hbox{\input{symbols/DdotSym.tex}}\!\!_K$ respectively (we used balanced-symmetry to keep our diagrams tidy):
\begin{equation}
\resizebox{\textwidth}{!}{\input{pictures/adjoinConditionsImplication.tikz}}
\end{equation}
We then use the inverse condition for $F$ together with the fact that the antipode is self-inverse to obtain the transpose condition for $F^\dagger$.
\qed

Before proceeding to our main result, we wish to highlight the connection between strong complementarity, the key ingredient in our proof, and the quantum Fourier transform, the key ingredient in the traditional proof \cite{Jozsa1997,Jozsa2001}. 
\begin{lem}
Let $(\hbox{\input{symbols/ZbwdotSym.tex}}\!\!,\hbox{\input{symbols/DdotSym.tex}}\!\!)$ be a strongly complementary pair on some object $\SpaceH$ of some $\dagger$-SMC. Then the adjoints of the $\hbox{\input{symbols/DdotSym.tex}}\!\!$-classical states satisfy the following equations:
\begin{equation}\label{eqns_multiplicativeChar}
\input{pictures/multiplicativeChar.tikz}
\end{equation}
We will refer to them as the $(\hbox{\input{symbols/ZbwdotSym.tex}}\!\!,\hbox{\input{symbols/DdotSym.tex}}\!\!)$-multiplicative characters. 
\end{lem}
\proof
The first two equations are the daggers of the copy and delete conditions for $\hbox{\input{symbols/DdotSym.tex}}\!\!$-classical states. The last equation follows by post-composing the dagger of the transpose condition with the symmetric cap for $\hbox{\input{symbols/ZbwdotSym.tex}}\!\!$. 
\qed
\begin{cor}
Let $(\hbox{\input{symbols/ZbwdotSym.tex}}\!\!,\hbox{\input{symbols/DdotSym.tex}}\!\!)$ be a strongly complementary pair on a finite-dimensional quantum system $\SpaceH$, with $\hbox{\input{symbols/ZbwdotSym.tex}}\!\!$ a $\dagger$-SCFA and $\hbox{\input{symbols/DdotSym.tex}}\!\!$ a $\dagger$-qSCFA. Let $G := (\classicalStates{\hbox{\input{symbols/ZbwdotSym.tex}}\!\!},\!\hbox{\input{symbols/DmultSym.tex}}\!\!,\!\hbox{\input{symbols/DunitSym.tex}}\!\!)$ be the finite abelian group corresponding to the pair, and $G^\wedge$ be its Pontryagin dual, i.e. the finite abelian group of multiplicative characters for $G$. If the non-degenerate observable corresponding to $\hbox{\input{symbols/ZbwdotSym.tex}}\!\!$ is taken as the computational basis $(\ket{g})_{g \in G}$, then the non-degenerate observable corresponding to $\hbox{\input{symbols/DdotSym.tex}}\!\!$ is the Fourier basis $(\ket{\goodchi})_{\goodchi \in G^\wedge}$ corresponding to the group $G$, i.e. the basis given by the following states:
\begin{equation}\label{FourierBasisExplicit}
\ket{\goodchi} = \sum_{g \in G} \goodchi(g) \ket{g}
\end{equation}
\end{cor}
\proof
Applying Equations \ref{eqns_multiplicativeChar} to the $\hbox{\input{symbols/ZbwdotSym.tex}}\!\!$-classical states $(\ket{g})_{g \in G}$ yields maps $\goodchi: G \rightarrow \complexs$ for all $\hbox{\input{symbols/DdotSym.tex}}\!\!$-classical states $\bra{\goodchi} \in \classicalStates{\hbox{\input{symbols/DdotSym.tex}}\!\!}$. These maps satisfy $\goodchi(gh) = \goodchi(g)\goodchi(h)$, $\goodchi(1) = 1$ and $\goodchi(g^{-1}) = \goodchi(g)^\dagger$, and hence are multiplicative characters. Conversely, the adjoint of any state $\ket{\goodchi}$ define as in Equation \ref{FourierBasisExplicit} for a multiplicative character $\goodchi$ satisfies Equations \ref{eqns_multiplicativeChar} when applied to the $\hbox{\input{symbols/ZbwdotSym.tex}}\!\!$-classical states, and hence satisfies them altogether because $\hbox{\input{symbols/ZbwdotSym.tex}}\!\!$ has enough classical states. Hence the $\hbox{\input{symbols/DdotSym.tex}}\!\!$-classical states are in bijection with the multiplicative characters of $G$.
\qed

\section{The Quantum Algorithm -- Traditional Presentation}
\label{section_algoTrad}

The quantum subroutine of the algorithm solving the abelian HSP proceeds as follows:
\begin{enumerate}
	\item[(i)] an initial state $\sum_{g \in G}\ket{g} \otimes \ket{0}$ of the joint quantum system $\complexs[G] \otimes \complexs[\integersMod{2}^N]$ is prepared;
	\item[(ii)] the unitary oracle $U_f$ is applied to this initial state;
	\item[(iii)] the subsystem $\complexs[\integersMod{2}^N]$ is measured in the standard orthonormal basis, resulting in a classical measurement outcome $b \in \integersMod{2}^N$;
	\item[(iv)] the subsystem $\complexs[G]$ is measured in the orthonormal basis $\big(\frac{1}{\sqrt{|G|}} \, \sum_{g \in G} \goodchi(g) \ket{g} \big)_{\goodchi \in G^\wedge}$, where $G^\wedge$ is the set of multiplicative characters\footnote{In fact, multiplicative characters $\goodchi: G \rightarrow S^1$ form a finite abelian group with pointwise multiplication.} of the finite abelian\footnote{The requirement that $G$ be abelian is important here: the linear extensions of multiplicative characters form a basis of the group algebra $\complexs[G]$ if and only if $G$ is abelian.} group $G$.
\end{enumerate} 
Now assume that the measurement on $\complexs[\integersMod{2}^N]$ has yielded a string $b$, associated with a coset $g_b H$ of the hidden subgroup $H$ (i.e. an element of the quotient group $G/H$). Then we can compute the probability of an outcome $\goodchi \in G^\wedge$ (where $G^\wedge$ denotes the abelian group of multiplicative characters of $G$): they main observation behind the algorithm is that this results in a uniform distribution over the \textbf{annihilator} of $H$, the subgroup $\Annihil{H} \leq G$ containing those multiplicative characters such that $\goodchi(h) = 1$ for all $h \in H$ (independently of which $b$ we had obtained). Therefore, the quantum subroutine of the algorithm provides a way to uniformly sample the annihilator of the hidden subgroup $H$.

The classical part of the algorithm identifies $H$ by sampling its annihilator a number of times logarithmic in the size of $H$. Because $G$ is abelian, the \textit{Fundamental Theorem for finite abelian groups} \cite{Frobenius1878} implies that it admits a description in terms of $O(\log |G|)$ generators, i.e. that there are $k \propto \log |G|$ generators $g_1,...,g_k$ such that every $g \in G$ can be written uniquely as $g_1^{n_1} \cdot ... \cdot g_k^{n_k}$, for coefficients $n_j \in \integersMod{\operatorname{ord}(g_j)}$. In terms of generators, the multiplicative characters take the following form, for all possible $p := (p_1,...,p_k) \in \prod_{j=1}^k \integersMod{\operatorname{ord}(g_j)}$: 
\begin{equation}
\goodchi_{p}\big(g_1^{n_1} \cdot ... \cdot g_k^{n_k}\big) = e^{i 2 \pi \, \big( \frac{p_1 \cdot n_1}{\operatorname{ord}(g_1)} + ... + \frac{p_k \cdot n_k}{\operatorname{ord}(g_k)} \big)}
\end{equation}
Each sampled $\goodchi_p \in \Annihil{H}$ then results in the following equation (coming from $\goodchi_p(h) = 1$), which all elements $h := g_i^{m_1}\cdot...\cdot g_k^{m_k}$ of the hidden subgroup $H$ must satisfy:
\begin{equation}
\frac{p_1 \cdot m_1}{\operatorname{ord}(g_1)} + ... + \frac{p_k \cdot m_k}{\operatorname{ord}(g_k)} = \modclass{0}{1} 
\end{equation}
Because $H$ is an abelian subgroup, it also admits a logarithmic description in terms of the generators, which can be identified by simultaneously solving $O(\log |H|)$ such equations.

\section{The Quantum Algorithm -- Diagrammatic Presentation}
\label{section_algoDiag}

For the fully diagrammatic version of the proof, we replace the concrete Hilbert space setting with four assumptions about the $\dagger$-SMC we want to implement the protocol in, and the strongly complementary pairs that it possesses.

\begin{enumerate}
	\item[(a)] There exist strongly complementary pairs encoding the four relevant finite abelian groups: the group $G$, the hidden subgroup $H$, the quotient group $G/H$ and the group of $N$-bit strings $\integersMod{2}^N$. That is, there exists a strongly complementary pair $(\hbox{\input{symbols/ZbwdotSym.tex}}\!\!_K,\hbox{\input{symbols/DdotSym.tex}}\!\!_K)$ on object $\SpaceH_K$ such that $K \isom (\classicalStates{\hbox{\input{symbols/ZbwdotSym.tex}}\!\!_K},\!\hbox{\input{symbols/DmultSym.tex}}\!\!_K,\!\hbox{\input{symbols/DunitSym.tex}}\!\!_K)$, for each $K=G,H,G/H,\integersMod{2}^N$.
	\item[(b)] The $\dagger$-SMC we are working with has an absorbing scalar $0$, i.e. we can define a sensible notion of impossibility in it.
	\item[(c)] $\hbox{\input{symbols/ZbwdotSym.tex}}\!\!_{H}$ and $\hbox{\input{symbols/ZbwdotSym.tex}}\!\!_{G/H}$ have enough classical states.
	\item[(d)] $\hbox{\input{symbols/DdotSym.tex}}\!\!_{G}$ and $\hbox{\input{symbols/ZbwdotSym.tex}}\!\!_{\integersMod{2}^N}$ have enough classical states, and that their classical states are orthogonal, so that measurement in either observable can be properly interpreted as a process with classical output.
\end{enumerate}
As a matter of convenience, we also assume the point structures $\hbox{\input{symbols/ZbwdotSym.tex}}\!\!_K$ to be special, though this is not crucial to the proof. We denote by $\xi_K$ the normalisation scalars of the groups structures $\hbox{\input{symbols/DdotSym.tex}}\!\!_K$ (which are quasi-special).
\begin{rmrk}
When working in the $\dagger$-SMC $\fdHilbCategory$ of finite-dimensional Hilbert spaces (which obviously satisfies condition (b)), condition (a) is guaranteed by the classification theorem for strongly complementary pairs of $\dagger$-SCFAs in $\fdHilbCategory$ \cite{Kissinger2012} (because the groups we consider are finite abelian). Conditions (c) and (d) hold automatically if we pick the $\hbox{\input{symbols/ZbwdotSym.tex}}\!\!_K$ algebras to be commutative \cite{Coecke2013b}.
\end{rmrk}
\vspace{-2mm}

\subsection{Constructing the oracle}

We begin by constructing an abstract version of the unitary oracle $U_f$ given in Equation \ref{eqn_unitaryOracle}. We replace the subgroup hiding function $f: G \rightarrow \integersMod{2}^N$ (or, to be precise, its linear extension $f: \complexs[G] \rightarrow \complexs[\integersMod{2}^N]$) with a $\hbox{\input{symbols/ZbwdotSym.tex}}\!\!_G$-to-$\hbox{\input{symbols/ZbwdotSym.tex}}\!\!_{\integersMod{2}^N}$-classical map $f: \SpaceH_G \rightarrow \SpaceH_{\integersMod{2}^N}$, which is required to satisfy an appropriate factorisation promise (detailed later on). The unitary oracle $U_f$ can then be decomposed as follows, in terms of a coherent copy operations for $G$, a coherent multiplication operations for $\integersMod{2}^N$, and the coherent subgroup hiding function $f$:
\begin{equation}\label{eqn_unitaryOracle1}
\input{pictures/unitaryOracle.tikz}
\end{equation}
The process $U_f$ defined above is always unitary, and on finite-dimensional quantum systems it coincides with the oracle we explicitly defined in the previous Section \cite{Vicary2012a}.

\subsection{What we want to show}

The following diagram presents the quantum subroutine in its entirety: the initial state is prepared, the unitary oracle is applied, and two outcomes $b \in \integersMod{2}^N$ and $\goodchi \in G^\wedge$ are obtained from the measurements performed on the two parts of the resulting state:
\begin{equation}\label{eqn_quantumHSPsubroutine}
\input{pictures/quantumHSPsubroutine.tikz}
\end{equation}
The diagram given above is a scalar $c_{b,\goodchi}$, and we interpret its square absolute value as the probability $\mathbb{P}(b,\goodchi) = c_{b,\goodchi}^\dagger c_{b,\goodchi}$ of obtaining the joint measurement outcome $(b,\goodchi)$. In the remainder of this Section, we will provide a fully diagrammatic proof that the probability must be zero if $b \notin \im{s}$ or if $\goodchi \notin  \Annihil{H}$, and it must otherwise be non-zero and independent of $b$ and $\goodchi$. In other words, we wish to show that the procedure produces a uniformly random sampling of the annihilator of $H$:
\begin{equation}
	\input{pictures/quantumHSPsubroutineValue.tikz}
\end{equation}
In the case of finite-dimensional quantum systems, we have $\xi_K^\dagger \xi_K = |K|$ for any finite group $K$, and hence the scalar appearing on the RHS is $|H|^2/|G|^2$. This is what we expect: there are $|G/H| = |G| / |H|$ distinct $b$ in the image of $s:G/H \rightarrow \integersMod{2}^N$ (because $s$ is injective), and the annihilator of $H$ itself has size $|G|/|H|$, leading to a total of $|G|^2/|H|^2$ possible joint measurement outcomes $(b,\goodchi)$.

\subsection{Using the Factorisation Promise}
We begin by decomposing $f$ according to the promise of factorisation through the quotient group $G/H$. Directly translated, Diagram \ref{diagram_quotientFactorisation} says that the coherent subgroup hiding function $f: \SpaceH_G \rightarrow \SpaceH_{\integersMod{2}^N}$ must factor through two maps $q: \SpaceH_{G} \rightarrow \SpaceH_{G/H}$ and $s: \SpaceH_{G/H} \rightarrow \SpaceH_{\integersMod{2}^N}$ 
\begin{equation}
\input{pictures/quotientFactorisationCoherent.tikz}
\end{equation}
The map $q$ corresponds to the quotient group homomorphism, and hence must be a $(\hbox{\input{symbols/ZbwdotSym.tex}}\!\!_G,\hbox{\input{symbols/DdotSym.tex}}\!\!_G)$-to-$(\hbox{\input{symbols/ZbwdotSym.tex}}\!\!_{G/H},\hbox{\input{symbols/DdotSym.tex}}\!\!_{G/H})$ homomorphism (satisfying three additional properties detailed later on), while the map $s$ must correspond to an injection, and hence must be a  $\hbox{\input{symbols/ZbwdotSym.tex}}\!\!_{G/H}$-to-$\hbox{\input{symbols/ZbwdotSym.tex}}\!\!_{\integersMod{2}^N}$-classical map and an isometry\footnote{When classical data is encoded on orthonormal bases, the coherent versions of injections are isometries.}. As our first manipulation step, we can substitute the promised factorisation of $f$ into $s \circ q$, and use the unit law to remove the $\dagger$-qSCFA $\hbox{\input{symbols/DdotSym.tex}}\!\!_{\integersMod{2}^N}$ from the diagram:
\begin{equation}\label{eqn_quantumHSPsubroutine1}
\input{pictures/quantumHSPsubroutine1.tikz}
\end{equation}

\subsection{Eliminating the $\integersMod{2}^N$ labels}

The property of process $s: \SpaceH_{G/H} \rightarrow \SpaceH_{\integersMod{2}^N}$ being an isometry can be readily formulated diagrammatically as follows:
\begin{equation}
\input{pictures/isometry.tikz}
\end{equation}
Because $s$ is a $\hbox{\input{symbols/ZbwdotSym.tex}}\!\!_{G/H}$-to-$\hbox{\input{symbols/ZbwdotSym.tex}}\!\!_{\integersMod{2}^N}$ classical map, and because $\hbox{\input{symbols/ZbwdotSym.tex}}\!\!_{G/H}$ has enough classical states and the classical states of $\hbox{\input{symbols/ZbwdotSym.tex}}\!\!_{\integersMod{2}^N}$ are orthogonal, then we have the following: a $\hbox{\input{symbols/ZbwdotSym.tex}}\!\!_{\integersMod{2}^N}$-classical state $b$ is either in the image of $s$, i.e. $b = s \circ (g_b H)$ for some classical state $g_b H$, or we have that $b^\dagger \circ s = 0$ is the impossible process, i.e. $b$ is never observed as outcome. Our second manipulation step then assumes that the outcome $b \in \integersMod{2}^N$ is in the image of $s$, and uses the isometry property to remove $s$ from the diagram altogether:
\begin{equation}\label{eqn_quantumHSPsubroutine2}
\input{pictures/quantumHSPsubroutine2.tikz}
\end{equation}

\subsection{Formalising the Quotient Map}

The manipulation of $q$ in the diagram is more complicated than that of $s$, since much of the proof relies on the fact that it is the coherent version of a very specific group homomorphism. Operationally, the quotient group homomorphism $q: G \rightarrow G/H$ can be characterised as one satisfying the following three properties:
\begin{enumerate}
	\item[(a)] $q$ identifies enough elements of $G$ to send all of $H$ to the group unit;
	\item[(b)] $q$ does not send any elements other than those in $H$ to the group unit;
	\item[(c)] $q$ is surjective.
\end{enumerate}
These three properties can be captured graphically as follows, where $r: \SpaceH_{G/H} \rightarrow \SpaceH_{G}$ is some isometry (a section for $q$, witnessing its surjectivity), and $i_H : \SpaceH_{H} \rightarrow \SpaceH_{G}$ is a $\hbox{\input{symbols/ZbwdotSym.tex}}\!\!_{H}$-to-$\hbox{\input{symbols/ZbwdotSym.tex}}\!\!_{G}$ classical isometry, corresponding to the group homomorphism injecting $H$ as a subgroup of $G$:
\begin{equation}\label{eqns_quotientMap}
\input{pictures/quotientMap.tikz}
\end{equation}
As a matter of notational convenience, we will henceforth choose representatives $g_b H$ in the quotient group so that $r \ket{g_b H} := \ket{g_b}$ holds.

\subsection{Excluding the Non-Annihilator Outcomes}

As our next step, we want to show that the diagram evaluates to $0$ whenever $\goodchi$ is not in the annihilator of $H$. To do so, we first need a graphical definition of what it means for a character $\goodchi$ to annihilate $H$:
\begin{equation}\label{eqn_annihilator}
\input{pictures/annihilator.tikz}
\end{equation}
We begin by moving $q$ from the lower to the upper branch, by using the fact that it is $\hbox{\input{symbols/ZbwdotSym.tex}}\!\!_{G}$-to-$\hbox{\input{symbols/ZbwdotSym.tex}}\!\!_{G/H}$ classical:
\begin{equation}\label{eqn_quantumHSPsubroutine3a}
\input{pictures/quantumHSPsubroutine3a.tikz}
\end{equation}
We then proceed to show that either $\goodchi^\dagger \circ q^\dagger = 0$, and hence that the entire diagram vanishes, or that the character $\goodchi$ is in the annihilator of $H$ (according to the graphical definition of Equation \ref{eqn_annihilator}): 
\begin{equation}\label{eqn_killingNonAnnihilator}
\input{pictures/killingNonAnnihilator.tikz}
\end{equation}
The first equality is a consequence of the following equivalent reformulation of a defining property of the quotient map $q$:
\begin{equation}\label{eqn_shortExactPropertyAlt}
\input{pictures/shortExactPropertyAlt.tikz}
\end{equation}
The equivalence between the two versions can be proven by using the same general technique which is used in Equation \ref{eqns_cosetStatesDecompositionProof2} below.

\subsection{Introducing the Coset States}

Having excluded the case where $\goodchi \notin \Annihil{H}$ (which won't really be needed until the next subsection), our third manipulation step goes as follows:
\begin{equation}\label{eqn_quantumHSPsubroutine3b}
\input{pictures/quantumHSPsubroutine3b.tikz}
\end{equation}
We removed both $q^\dagger$ and the state $g_b H$ of $\SpaceH_{G/H}$ from the diagram, and replaced them with an abstract version of the \textbf{coset state} $\sum_{h \in H} \ket{g_b \cdot h}$ of $\SpaceH_{G}$, by using the following result:
\begin{equation}\label{eqn_cosetStatesDecompositionStatement}
\input{pictures/cosetStatesDecompositionStatement.tikz}
\end{equation}
The equality above can be proven diagrammatically using the defining properties of $q$: 
\begin{equation}
\input{pictures/cosetStatesDecompositionProof1.tikz}
\end{equation}
More in detail, the second equation in the chain is proven by using the fact that $q$ is a $(\hbox{\input{symbols/ZbwdotSym.tex}}\!\!_G,\hbox{\input{symbols/DdotSym.tex}}\!\!_G)$-to-$(\hbox{\input{symbols/ZbwdotSym.tex}}\!\!_{G/H},\hbox{\input{symbols/DdotSym.tex}}\!\!_{G/H})$ homomorphism, by replacing the antipode with its definition, and by appealing to the fact that the antipode is self-inverse:
\begin{equation}\label{eqns_cosetStatesDecompositionProof2}
\resizebox{\textwidth}{!}{\input{pictures/cosetStatesDecompositionProof2.tikz}}
\end{equation}

\subsection{Annihilating the Coset States}

We are now in the situation where $\goodchi$ is in the annihilator of $H$, and we have rewritten our diagram explicitly in terms of coset states. As our fourth manipulation step, we turn the character around to obtain (the adjoint of) a diagram involving a character evaluated on a coset state:
\begin{equation}\label{eqn_quantumHSPsubroutine4}
\input{pictures/quantumHSPsubroutine4.tikz}
\end{equation}
Because the character is multiplicative (its adjoint is a $\hbox{\input{symbols/DdotSym.tex}}\!\!_G$-classical state), we can copy it through~$\!\hbox{\input{symbols/DcomultSym.tex}}\!\!_G$. Evaluating against $g_b^{-1}$ removes the first copy to give some phase $\goodchi(g)$, satisfying $\goodchi(g)^\dagger \goodchi(g) = 1$, while the definition of the annihilator removes the second copy together with $i_H^\dagger$:
\begin{equation}\label{eqn_quantumHSPsubroutine5}
\input{pictures/quantumHSPsubroutine5.tikz}
\end{equation}
We are left with a bunch of explicit scalars, and we can finally evaluate the square absolute value of Diagram \ref{eqn_quantumHSPsubroutine} to obtain our desired result: 
\begin{equation}
\resizebox{\textwidth}{!}{\input{pictures/quantumHSPsubroutineValueFinal.tikz}}
\end{equation}
The evaluation of the square norm $\!\hbox{\input{symbols/ZbwunitSym.tex}}\!\!_H^\dagger\circ\!\hbox{\input{symbols/ZbwunitSym.tex}}\!\!_H $ to $\xi_H^\dagger \xi_H$ comes from the fact that $\!\hbox{\input{symbols/ZbwunitSym.tex}}\!\!_H$ is $\hbox{\input{symbols/DdotSym.tex}}\!\!_H$-classical, and the latter has $\xi_H$ as normalisation factor.

\section{Non-abelian HSP}
\label{section_nonAbelian}

Nowhere in the proof above we have explicitly used commutativity of the $\dagger$-qSCFAs (equivalently, the fact that $G$ and $H$ are abelian), and our approach naturally generalises to the case where $G$ is a finite group and $H$ a normal subgroup (a necessary requirement in this approach, which explicitly uses a group structure on $G/H$). For the sake of simplicity, and because no hard result will be proven, we will stick to the case of finite-dimensional Hilbert spaces for the remainder of this Section. 

Going from commutative to general quasi-special $\dagger$-Frobenius algebras ($\dagger$-qSFA) has the following implication: the classical states are still the multiplicative characters, and they are still orthogonal, but they no longer form a basis. Instead, the $\dagger$-qSFA is now associated with a potentially degenerate observable: sampling it will produce, as classical output, the character $\goodchi_\rho$ of an irreducible representation $\rho$ of $G$, with the following probability (where $d_\rho$ is the dimension of representation $\rho$):
\begin{equation}
	\mathbb{P}[b,\goodchi_\rho] = 
	\begin{cases}
		\frac{|H|^2}{|G|^2} d_\rho^2  & \text{ if } \rho(h) = 0 \text{ for all } h \in H \\
		0 & \text{ otherwise }
	\end{cases}
\end{equation}
For our graphical proof to go through, Diagram \ref{eqn_quantumHSPsubroutine} needs to be modified as follows:
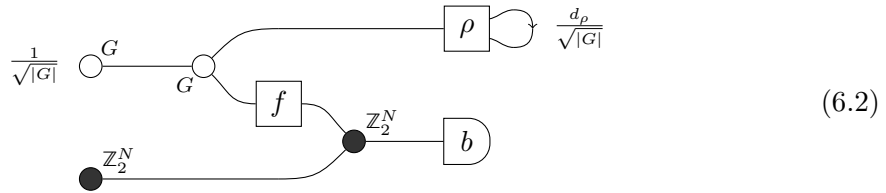
\begin{equation}\label{eqn_quantumHSPsubroutineNonAbelian}
\input{pictures/quantumHSPsubroutineNonAbelian.tikz}
\end{equation}
where we used the dagger-compact structure to take the trace of the irreducible representation $\rho: \complexs[G] \rightarrow V_\rho^\ast \otimes V_\rho$ (technically, its linear extension from $G$ to $\complexs[G]$).
The defining properties of multiplicative characters generalise to representations \cite{Vicary2012a}:
\begin{equation}\label{eqn_irrep}
\resizebox{\textwidth}{!}{\input{pictures/irrep.tikz}}
\end{equation} 
However, the generalisation from the abelian to the non-abelian case encounters a much bigger hurdle in the classical post-processing: the logarithmic dependency of the number of generators on the size of the group only need to hold in the abelian case, and the number of samples required is in general linear in the size of the group (and hence exponential in the size of its description) \cite{Hallgren2010}. This is a separate problem, interesting in its own right, and is beyond the scope of this work.

\section{Further Applications}

The fully diagrammatic, abstract character of our approach means that our results can be directly applied to theories other than quantum theory, as long as they feature the relevant algebraic structure. We give three short examples, to corroborate our point: Real Quantum Theory; more general toy models of quantum theory based on semirings other than $\complexs$ and $\reals$; and treatment of infinite-dimensional HSP using non-standard analysis.

\subsection{Simon's Problem in Real Quantum Theory}
\label{section_SimonsProblemRealQT}

While the quantum subroutine for the HSP can be formulated with any theory satisfying the few technical requirements we have imposed, the multiplicative characters appearing in the theory may be very different from the complex-valued ones appearing in the traditional implementation, and this may give rise to issues in the classical post-processing part of the algorithm (e.g. it might be hard/impossible to reconstruct the hidden subgroup from the characters sampled by the routine). As a concrete example, we consider the case of Real Quantum Theory \cite{Belenchia2012}, where wavefunctions are restricted to take real-valued amplitudes only. Real quantum theory is modelled by finite-dimensional real Hilbert spaces, and satisfies all the requirements for the quantum subroutine to be implemented (for all finite abelian groups $G$ and all subgroup hiding functions). However, the multiplicative characters arising in Real Quantum Theory are only the real-valued characters of $G$: this means that, except in the case $G = \integersMod{2}^M$, the classical post-processing part of the algorithm will fail in most cases. Thus said, it is interesting to note that both the quantum subroutine and the classical post-processing will work out when $G = \integersMod{2}^M$. As a consequence, Simon's problem can be efficiently solved in Real Quantum Theory, and the class BQP$_\reals$ (by which we mean BQP for Real Quantum Theory) is separated from BPP relative to oracles with the appropriate promise.

\subsection{Categories of modules over semirings}
\label{section_categoriesOfSemirings}

Spekkens' toy model \cite{Spekkens2007,Catani2017} is a toy theory, based on sets and relations, which shares many of the iconic operational features of quantum theory while at the same time being fully local. The toy model has been formulated categorically and diagrammatically \cite{Coecke2012a,Backens2015}, and it has been used, in comparison with the ZX calculus for qubit stabiliser quantum theory, to investigate the relationship between non-locality and phase groups \cite{Coecke2010a}. More recently, formulations of several quantum algorithms and protocols have been studied in terms of finite sets and relations, as opposed to finite-dimensional Hilbert spaces, and more specifically within Spekkens' toy model \cite{Zeng2015,Disilvestro2016}. In particular, a first translation of the HSP quantum algorithm to the category of sets and relation was given in \cite{Zeng2015}. 

We take here a broader outlook on things, and consider the extremely general case of categories of modules over involutive commutative semirings, of which real/complex quantum theory and the category of sets and relations are all special cases (for the semirings of the real/complex numbers and the semiring of the booleans respectively). Our approach, using strong complementarity and diagrammatic methods, will give us control over the HSP in this broader context, where the Fourier transform itself has not yet been studied. To be precise, we will fix an arbitrary commutative semiring $R$ with a (possibly trivial) involution $\dagger$ (that is, a self-inverse semiring homomorphism $\dagger: R \rightarrow R$), and we define the $\dagger$-SMC $\RMatCategory{R}$ of free, finite-dimensional modules over $R$:
\begin{enumerate}
	\item[(i)] the objects of $\RMatCategory{R}$ are labelled by finite sets;
	\item[(ii)] the morphisms $X \rightarrow Y$ in $\RMatCategory{R}$ are the $R$-valued matrices indexed by $Y \times X$, i.e. the free, finite-dimensional $R$-module $R^{Y \otimes X}$;
	\item[(iii)] composition and tensor product are the usual ones for matrices, while the dagger $f^\dagger: Y \rightarrow X$ of a morphism $f: X \rightarrow Y$ is given by taking the transpose of the matrix and then applying the involution to all elements (i.e. $(f^\dagger)_{yx} := (f_{xy})^\dagger$).
\end{enumerate} 
The category $\RMatCategory{R}$ is enriched over itself, and in particular we have a summation operation $+$ (matrix addition) and zero morphisms $0$ (the zero matrix).

\noindent Our reason to consider $\RMatCategory{R}$ is the following: on each object $X$ we can define a canonical copy/delete coalgebra by Equations \ref{coherentCopyDelete} (with associated $\dagger$-SCFA), which has the points of the finite set $X$ as classical states. Furthermore, for any choice of group structure $G := (X,\cdot,1)$ on $X$ we can define a group algebra by Equations \ref{groupAlgebra}: as long as the scalar $|G|$ (which is well defined in any semiring) can be factored as $|G| = \xi_G^\dagger\xi_G$, the group algebra always gives rise to a $\dagger$-qSFA with normalisation factor $\xi_G$. If we are willing to weaken the definition of quasi-special $\dagger$-Frobenius algebra to include any scalar $\Xi$, instead of only positive ones $\Xi := \xi^\dagger \xi$, then the group algebra always gives rise to a $\dagger$-qSFA this way: our proof does not require positivity, and goes through with minor modification under this more general definition (but we lost the equivalence between quasi-special and special $\dagger$-Frobenius algebras in the process, which creates problems with our operational interpretation).

The change from the complex numbers to an arbitrary semiring affects the structure of characters and representations of groups, with consequent impact on the measurement in $\hbox{\input{symbols/DdotSym.tex}}\!\!_G$ and on the classical post-processing. Firstly, even in the abelian case there may not be enough multiplicative characters $G \rightarrow (R^\times, \cdot, 1)$ to distinguish the elements of $G$ themselves, so that $\hbox{\input{symbols/DdotSym.tex}}\!\!_G$ fails to have enough classical states: this is usually the case in the category of sets and relations, because when $R$ is the semiring of the booleans the only character available is the trivial character. Secondly, even if there are enough multiplicative characters to distinguish group elements, the classical post-processing step is not guaranteed to go through on the nose. A detailed study of the structure of characters (i.e. of Fourier theory over involutive commutative semirings) and of the required generalisation of the classical post-processing is left to future work.

\subsection{Infinite-dimensional HSP}
\label{section_infiniteHSP}

The category $\HilbCategory$ of infinite-dimensional Hilbert spaces and bounded linear maps does not admit Frobenius algebras, and as a consequence it cannot be used to extend our abstract setup to infinite abelian groups. However, tools from non-standard analysis can be used to construct a well-defined category $\starHilbCategory$ of infinite-dimensional separable Hilbert spaces, including both bounded and unbounded linear maps, as well as a number of commonplace features of quantum mechanics (such as Dirac deltas and plane-waves) \cite{Gogioso2017}. 

It has been shown that the category $\starHilbCategory$ possesses suitable strongly complementary pairs corresponding to the discrete groups $\integers^N$ of translations of lattices and the compact groups $T^N$ of translations of tori; all the observables concerned are quasi-special or special, commutative, and have enough classical states. These observables have direct physical relevance, as they correspond to the momentum/position observable pairs for particles in $N$-dimensional boxes with periodic boundary conditions. As a consequence, our scheme straightforwardly extends to a quantum subroutine for the HSP on the infinite abelian groups $G = \integers^N$ (the translation groups of lattices) or their Pontryagin duals $G = T^N$ (the translation groups of tori). 

The classical subroutine requires no adjustment for the $G = \integers^N$ cases: all the possible quotients $G/H$, and hence all the possible annihilators $\Annihil{H} \leq T^N$, are finite. For the sake of physical implementation, this corresponds to fixing the computational basis $\hbox{\input{symbols/ZbwdotSym.tex}}\!\!_G$ in the momentum eigenstates (valued in $\integers^N$) of a particle in an $N$-dimensional box with periodic boundary conditions, and performing the $\hbox{\input{symbols/DdotSym.tex}}\!\!_G$ measurement corresponding to its position observable (valued in $T^N$). A similar physical setup, with position and momentum swapped, can be used to tackle the $G = T^N$ case, but some minor adjustment will be required in the classical subroutine (because the annihilators $\Annihil{H} \leq \integers^N$ are all infinite).

\section{Conclusions}
\label{section_conclusions}

The abelian Hidden Subgroup Problem comprises many of the problems successfully tackled by quantum algorithms as special instances, but the traditional presentation of the quantum solution is too heavily algebraic to clearly show the key structures at work. We improved upon previous work by presenting the first fully graphical proof of correctness for the algorithm, showing that strong complementarity is the key algebraic feature behind the quantum advantage in the abelian HSP.

We have remarked that our diagrammatic treatment naturally extends to the non-abelian case, and that the known intractability of the problem is more a matter of classical post-processing than an issue with the quantum part itself. We have also shown that our approach immediately applies to theories other than finite-dimensional quantum theory, as long as they possess the required algebraic structures. We have shown that Simon's Problem can be efficiently solved in Real Quantum Theory, and we have highlighted the issues and opportunities associated with choices of semirings of scalars other than the traditional $\complexs$ (Quantum Theory) and $\reals$ (Real Quantum Theory). We have also seen that our fully diagrammatic approach immediately applies to the infinite-dimensional case of wavefunctions in a box with periodic boundary conditions, where it yields an algorithm to solve the HSP for the infinite abelian groups $\integers^N$ and $T^N$ (translations of $N$-dimensional lattices and tori, respectively).

A number of questions remain open. Firstly, the group theoretic nature of the Hidden Subgroup Problem begs the question of whether strong complementarity is somehow also a necessary condition for the implementation of a suitable quantum subroutine. Secondly, it would be interesting to look at more explicit implementations of our results in other suitable theories, such as Fermionic Quantum Theory \cite{DAriano2014}, Spekkens' Toy Model, and infinite-dimensional quantum theory. Finally, the relationship between strong complementarity and the quantum Fourier transform prompts further investigation of the role that these algebraic structures might be playing in a number of other quantum algorithms and protocols.

\section*{Acknowledgements.}
The authors would like to thank Bob Coecke for comments and suggestions, as well as Sukrita Chatterji and Nicol\`o Chiappori for support. The first author would like to acknowledge support from EPSRC (OUCL/2013/SG) and Trinity College (Williams Scholarship). The second author would like to acknowledge support from the ERC under the European Union's Seventh Framework Programme (FP7/2007-2013) / ERC grant n\textsuperscript{o} 320571.

\input{biblio.bbl}
\end{document}

%% file: packages.tex
\usepackage{mathtools} 
\usepackage{amssymb} 
\usepackage{bbold} 
\usepackage{amsthm} 

\usepackage{microtype} 
\usepackage{ragged2e}

\usepackage[nocompress]{cite} 

\usepackage{graphicx} 
\usepackage[usenames,dvipsnames]{xcolor} 
\usepackage{tikz} 
\usepackage{circuitikz} 
\usetikzlibrary{
	arrows,
	shapes,
	decorations,
	intersections,
	backgrounds,
	positioning,
	circuits.ee.IEC
}

%% file: macros.tex
\tikzstyle{empty diagram}=[draw=gray!40!white,dashed,shape=rectangle,minimum width=1cm,minimum height=1cm]

	\newcommand{\goodchi}{\protect\raisebox{2pt}{$\chi$}} 
	\newcommand{\verteq}{\rotatebox{90}{$\,=$}}

	\newcommand{\integers}{\mathbb{Z}} 
	\newcommand{\reals}{\mathbb{R}} 
	\newcommand{\complexs}{\mathbb{C}} 
	\newcommand{\integersMod}[1]{\mathbb{Z}_{#1}} 
	\newcommand{\modclass}[2]{#1 \; (\text{mod } #2)} 
	
		\newcommand{\ket}[1]{\vert #1 \rangle} 
		\newcommand{\bra}[1]{\langle #1 \vert} 
		
		\newcommand{\im}[1]{\operatorname{im}#1} 
		\newcommand{\Annihil}[1]{\operatorname{Ann}[#1]} 

		\newcommand{\SpaceH}{\mathcal{H}} 
		
		\newcommand{\SpaceK}{\mathcal{K}}

		\newcommand{\isom}{\cong} 

		\newcommand{\UnitaryOps}[1]{U\left[#1\right]} 

		\newcommand{\RMatCategory}[1]{#1\operatorname{-Mat}} 

		\newcommand{\HilbCategory}{\operatorname{Hilb}} 
		\newcommand{\fdHilbCategory}{\operatorname{fdHilb}} 
		\newcommand{\starHilbCategory}{^\star\!\HilbCategory} 

	\newcommand{\classicalStates}[1]{\mathfrak{K}(#1)} 

	\newcommand{\Xcolour}{Red}
	
	\newcommand{\Zcolour}{YellowGreen}

	\newcommand{\Xaltcolour}{Purple}
	
	\newcommand{\Zaltcolour}{Cyan}
	
	\newcommand{\Dcolour}{black!80}

	\newcommand{\Xbwcolour}{black!80}

	\newcommand{\Zbwcolour}{white}

	\newcommand{\Ybwcolour}{black!15}

	\newcommand{\Wbwcolour}{black!55}

	\tikzset{
	  rectangle with rounded corners north west/.initial=4pt,
	  rectangle with rounded corners south west/.initial=4pt,
	  rectangle with rounded corners north east/.initial=4pt,
	  rectangle with rounded corners south east/.initial=4pt,
	}
	\makeatletter
	\pgfdeclareshape{rectangle with rounded corners}{
	  \inheritsavedanchors[from=rectangle] 
	  \inheritanchorborder[from=rectangle]
	  \inheritanchor[from=rectangle]{center}
	  \inheritanchor[from=rectangle]{north}
	  \inheritanchor[from=rectangle]{south}
	  \inheritanchor[from=rectangle]{west}
	  \inheritanchor[from=rectangle]{east}
	  \inheritanchor[from=rectangle]{north east}
	  \inheritanchor[from=rectangle]{south east}
	  \inheritanchor[from=rectangle]{north west}
	  \inheritanchor[from=rectangle]{south west}
	  \backgroundpath{
	    \southwest \pgf@xa=\pgf@x \pgf@ya=\pgf@y
	    \northeast \pgf@xb=\pgf@x \pgf@yb=\pgf@y
	    \pgfkeysgetvalue{/tikz/rectangle with rounded corners north west}{\pgf@rectc}
	    \pgfsetcornersarced{\pgfpoint{\pgf@rectc}{\pgf@rectc}}
	    \pgfpathmoveto{\pgfpoint{\pgf@xa}{\pgf@ya}}
	    \pgfpathlineto{\pgfpoint{\pgf@xa}{\pgf@yb}}
	    \pgfkeysgetvalue{/tikz/rectangle with rounded corners north east}{\pgf@rectc}
	    \pgfsetcornersarced{\pgfpoint{\pgf@rectc}{\pgf@rectc}}
	    \pgfpathlineto{\pgfpoint{\pgf@xb}{\pgf@yb}}
	    \pgfkeysgetvalue{/tikz/rectangle with rounded corners south east}{\pgf@rectc}
	    \pgfsetcornersarced{\pgfpoint{\pgf@rectc}{\pgf@rectc}}
	    \pgfpathlineto{\pgfpoint{\pgf@xb}{\pgf@ya}}
	    \pgfkeysgetvalue{/tikz/rectangle with rounded corners south west}{\pgf@rectc}
	    \pgfsetcornersarced{\pgfpoint{\pgf@rectc}{\pgf@rectc}}
	    \pgfpathclose
	 }
	}
	\makeatother

	\tikzset{->-/.style={decoration={markings,mark=at position #1 with {\arrow{>}}},postaction={decorate}}}
	\tikzset{-<-/.style={decoration={markings,mark=at position #1 with {\arrow{<}}},postaction={decorate}}}

	\tikzstyle{every picture}=[baseline=-0.25em,scale=0.5]
	\pgfdeclarelayer{edgelayer}
	\pgfdeclarelayer{nodelayer}
	\pgfsetlayers{edgelayer,nodelayer,main}

	\tikzstyle{box} = [draw,shape=rectangle,inner sep=2pt,minimum height=6mm,minimum width=6mm,fill=white] 
	\tikzstyle{boxlarge} = [draw,shape=rectangle,inner sep=2pt,minimum height=1.5cm,minimum width=8mm,fill=white] 
	\tikzstyle{boxLarge} = [draw,shape=rectangle,inner sep=2pt,minimum height=2cm,minimum width=10mm,fill=white] 
	\tikzstyle{boxsmall} = [draw,shape=rectangle,inner sep=2pt,minimum height=3mm,minimum width=3mm,fill=white] 
	\tikzstyle{dot} = [inner sep=0mm,minimum width=3mm,minimum height=3mm,draw,shape=circle,text depth=-0.1mm]
	\tikzstyle{Zbwdot} = [dot, fill=\Zbwcolour]
	\tikzstyle{Xbwdot} = [dot, fill=\Xbwcolour]
	\tikzstyle{Ybwdot} = [dot, fill=\Ybwcolour]
	\tikzstyle{Wbwdot} = [dot, fill=\Wbwcolour]
	\tikzstyle{antipode} = [boxsmall] 

	\tikzstyle{state} = [draw, rectangle with rounded corners,
	  rectangle with rounded corners north west=8pt,
	  rectangle with rounded corners south west=8pt,
	  rectangle with rounded corners north east=0pt,
	  rectangle with rounded corners south east=0pt,
	,inner sep=2pt,minimum height=6mm,minimum width=6mm,fill=white]
	\tikzstyle{statelarge} = [draw, rectangle with rounded corners,
	  rectangle with rounded corners north west=8pt,
	  rectangle with rounded corners south west=8pt,
	  rectangle with rounded corners north east=0pt,
	  rectangle with rounded corners south east=0pt,
	,inner sep=2pt,minimum height=1.5cm,minimum width=8mm,fill=white]
	\tikzstyle{stateLarge} = [draw, rectangle with rounded corners,
	  rectangle with rounded corners north west=8pt,
	  rectangle with rounded corners south west=8pt,
	  rectangle with rounded corners north east=0pt,
	  rectangle with rounded corners south east=0pt,
	,inner sep=2pt,minimum height=2cm,minimum width=8mm,fill=white]
	\tikzstyle{effect} = [draw, rectangle with rounded corners,
	  rectangle with rounded corners north west=0pt,
	  rectangle with rounded corners south west=0pt,
	  rectangle with rounded corners north east=8pt,
	  rectangle with rounded corners south east=8pt,
	,inner sep=2pt,minimum height=6mm,minimum width=6mm,fill=white]
	\tikzstyle{scalar}=[diamond,draw,inner sep=1pt,font=\small,fill=white]

	\tikzstyle{cdnode}=[fill=white]
	\tikzstyle{labelnode}=[fill=white]
	\tikzstyle{tightlabelnode}=[fill=white,inner sep = 0.1mm]
	\tikzstyle{none}=[inner sep=0pt]
	\tikzstyle{whiteline}=[-, line width=4pt, draw=white]

	\tikzstyle{trace}=[circuit ee IEC,thick,ground,scale=2.5]
	\tikzstyle{cotrace}=[circuit ee IEC,thick,ground,rotate=180,scale=2.5]
	\tikzstyle{upground}=[circuit ee IEC,thick,ground,rotate=90,scale=2.5]
	\tikzstyle{downground}=[circuit ee IEC,thick,ground,rotate=-90,scale=2.5]

	\tikzstyle{doubled} = [line width=1.8pt] 